\documentclass[a4paper]{article}
\usepackage{a4wide}
\usepackage{amsmath,amssymb,amsfonts}
\usepackage[dvipdfm]{graphicx}
\usepackage{textcomp}
\usepackage{xcolor}
\def\BibTeX{{\rm B\kern-.05em{\sc i\kern-.025em b}\kern-.08em
    T\kern-.1667em\lower.7ex\hbox{E}\kern-.125emX}}
\usepackage{listings}
\usepackage{lmodern}
\begin{document}

\title{Accelerated Multiple Precision Direct Method and Mixed Precision Iterative Refinement on Python Programming Environment}
\author{Tomonori Kouya\\Shizuoka Institute of Science and Technology\\{https://na-inet.jp/}}
\date{2021-07-27(Tue)}
\maketitle

\begin{abstract}
Current Python programming environment does not have any reliable and efficient multiple precision floating-point (MPF) arithmetic except ``mpmath" and ``gmpy2" packages based on GNU MP(GMP) and MPFR libraries. Although it is well known that multi-component-type MPF library can be utilized for middle length precision arithmetic under 200 bits, they are not widely used on Python environment. In this paper, we describe our accelerated MPF direct method with AVX2 techniques and its application to mixed precision iterative refinement  combined with mpmath, and demonstrate their efficiency on x86\_64 computational environments.
\end{abstract}

%
\section{INTRODUCTION}

Python and its eco-system including tools of scientific computation, deep learning, and so on, are widely used all over the world. The Python eco-system is constructed on various modules and packages such as NumPy and SciPy. They are required enough to be efficient in order to be satisfied with users, so most popular Python modules are built on compiled C or C++ codes.

Current Python environment has only ``mpmath" and ``gmpy2" packages as efficient and rich multiple precision floating-point (MPF) arithmetic libraries. The mpmath package has well-known numerical functionalities such as basic linear algebra, but less efficient than gmpy2. On the other hand, gmpy2 has the same level of efficiency as using the native GMP and MPFR libraries directly, but it does not have the numerical computation capabilities provided by mpmath.

Although MPF packages such as mpmath and gmpy2 can support arbitrary precision floating-point arithmetic, when users require comparatively lower precision MPF from binary128 to binary256, multi-component-type MPF implementation, combined with two to four binary32 or binary64 hard-wired floating-point numbers, is appropriate. DD(Double-double) precision floating-point arithmetic is constructed with two binary64 numbers, and QD(Quad-double) precision floating-point arithmetic with four binary64 numbers. It is well-known that DD and QD precision arithmetics, which is originally implemented in QD library by Bailey et.al., are more efficient than MPF arithmetic of GMP and MPFR. On Python environment, QD-like multi-component-type MPF package is necessary to accelerate solvers of linear system of equations. In addition, implementing mixed precision iterative refinement with QD-like MPF package and mpmath can provide the best performance on Python environment.

We have already had BNCmatmul, accelerated MPF linear computation library based on DD, TD (Triple-double), and QD precision arithmetic\cite{kouya_iccsa2020}. It has been accelerated with AVX2 (Advanced Vector eXtension 2) available on x86\_64 CPUs, and has had also SIMDized direct method including LU decomposition and forward and backward substitutions\cite{kouya_arith28}. In this time, we have ported BNCmatmul to Python environment via DLL (dynamic linkage library) on Linux, and developed BNCamPy (Base for Numerical Computation with Accelerated Multiple Precision arithmetic on Python) package of Python to use the DLL. In this paper, we explain the construction of SIMDized MPF direct methods, and demonstrate their efficiency through the results of benchmark tests.

%
\section{IMPLEMENTATION OF MIXED PRECISION ITERATIVE REFINEMENT}

We have already confirmed that on Python environment, DD, TD, and QD precision MPF arithmetic are faster than the same precision MPF arithmetic of gmpy2 based on GMP and MPFR. And we have successfully implemented accelerated BNCmatmul library including LU decomposition.

\begin{figure}[htbp]
\centering{\includegraphics[width=.41\textwidth]{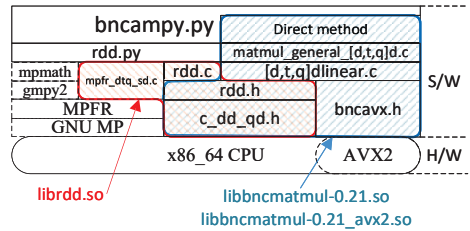}}
\caption{Software Layer of BNCmatmul and BNCamPy}\label{fig:lu_bench_layer}
\end{figure}

The software layer of our BNCmatmul and Python modules including direct method is shown in \figurename\ \ref{fig:lu_bench_layer}.

Basic DD, TD, and QD precision arithmetics are defined as C inline functions in c\_dd\_qd.h and also defined as macros in rdd.h. For usage of Python environment, the rdd.c, which is based on rdd.h, is compiled as librdd.so. DD, TD, and QD classes on Python utilize functions defined in librdd.so.

BNCmatmul, which is also built on rdd.h and c\_dd\_qd.h, has block and Strassen matrix multiplications that are better performed than simple triple-loop matrix multiplication. They have been accelerated using SIMDized functions with AVX2 defined in bncavx2.h. LU decomposition have been also accelerated with AVX2.

These function defined in BNCmatmul (\figurename\ \ref{fig:lu_bench_layer}) can be utilized from libbncmatmul-0.21.so without any SIMDization techniques and libbncmatmul-0.21\_avx2.so with SIMDization of AVX2. These two DLLs have the same name of functions, so users do not need to modify their codes when selecting these BNCmatmul DLLs.

The following Python code (\lstlistingname\ \ref{list:iterative_ref_td_mp.py}) is a mixed precision iterative refinement based on the algorithm originally proposed by Buttari et.al\cite{mixed_prec_iterative_ref}. It is a implementation of TD-mpmath type iterative refinement, and includes TD precision direct method (TDLUdecompPM and SolveTDLSPM functions of BNCmatmul) to obtain short precision approximation of solution.

{\small \begin{lstlisting}[caption=Main loop of TD-mpmath mixed precision iterative refinement method, label=list:iterative_ref_td_mp.py]
# Solve A * x = b where known A in RxR and b in R^n, unknown x in R^n
def iterative_refinement_td_mp(a, b, maxtimes, rtol, atol):
	dim = b.rows # set the number of dimension of linear system of equations

	# Initialize TD vectors and TD matrix
	af_ch = (ct.c_long * (dim))()
	af = init_tdmatrix(dim, dim);	bf = init_tdvector(dim);
	xf = init_tdvector(dim);	    x = init_mpvector(dim);
	res = init_mpvector(dim);	    resf = init_tdvector(dim);
	z = init_mpvector(dim);	      zf = init_tdvector(dim);

	# norm_a := ||A||_F
	norm_a = mpmath.mnorm(a, 'fro')

	# Make short precision copy of A and b
	af = get_tdmatrix_mpmat(af, a); bf = get_tdvector_mpvec(bf, b);

	# Compute short precision LU factorization
	# LU decomposition with partial pivoting and forward/backward substitutions
	TDLUdecompPM(af, af_ch); SolveTDLSPM(xf, af, bf, af_ch);

	# Port short precision solution to long precision one
	x = set_tdvector_mpvec(x, xf)

	# Repeat iterative refinement process
	for itimes in range(maxtimes):
		# Compute residual in long precision
		res = b - a * x

		# until ||r_i||_2 < sqrt(n) * reps * ||A||_F * ||x_i||_2
		norm_x = mpmath.norm(x); norm_res = mpmath.norm(res);

		if norm_res < mpmath.sqrt(mpmath.mpf(dim)) * rtol * norm_a * norm_x + atol:	break

		# Normalization: res := coef * res
		normalization_coef = mpmath.mpmathify(1) / norm_res
		res = normalization_coef * res

		# Port long precision residual to short precision one
		resf = get_tdvector_mpvec(resf, res)

		# Back-solve on short precision residual and short precision factors
		SolveTDLSPM(zf, af, resf, af_ch)

		# Port short precision correction to long precision one
		z = set_tdvector_mpvec(z, zf)

		# Reverse normalization
		z = norm_res * z

		# Update solution in long precision
		x = x + z
\end{lstlisting}}

%
\section{PERFORMANCE EVALUATION ON PYTHON ENVIRONMENT}

We have confirmed how our SIMDized direct methods of BNCmatmul library can be accelerated on two x86\_64 computational environments\cite{kouya_arith28}. The linear system of equations for our benchmark tests is 
\begin{equation}
	A\mathbf{x} = \mathbf{b}, \label{eqn:linear_eq}
\end{equation}
where $A\in\mathbb{R}^{n\times n}$, $\mathbf{x}\in\mathbb{R}^n$, and $\mathbf{b}\in\mathbb{R}^n$. $A$ and $\mathbf{b}$ are provided as follows:
\begin{description}
	\item[$A$] We produce $A := R D R^{-1}$ with appropriate precision mpmath, where the real diagonal matrix $D = \mathrm{diag}[d_1 \cdots d_n]$ ($d_i := 10^{-26(i-1)/n}$) and the real random matrix $R\in\mathbb{R}^{n\times n}$. Therefore $\kappa_2(A) = \|A\|_2\|A^{-1}\|_2 = 10^{26(n-1)/n}$.
	\item[$\mathbf{x}, \mathbf{b}$] The true solution is $\mathbf{x} = [0\ 1\ \cdots\ n-1]^T$, and we calculate $\mathbf{b} := A\mathbf{x}$ with mpmath.
\end{description}

The following x86\_64 computational environments are used through this paper.
\begin{description}\small
	\item[Corei9]:\vspace{1mm} Intel Core i9-10900X (3.6 GHz, 10 cores), 16 GB of RAM, Ubuntu 18.04.2, GCC 7.3.0, Python 3.8.3.
	\item[EPYC]:\vspace{1mm} AMD EPYC 7402P (1.5 GHz, 24 cores), 64 GB of RAM, Ubuntu 18.04.5, GCC 7.5.0, Python 3.8.3.
\end{description}

\tablename\ \ref{table:lubench} shows the results of benchmark test by running DD, TD, and QD precision direct methods. It includes computational times (Time(s)) and maximum relative errors of elements of numerical solutions (Max.RE) for all kinds of precisions and their accelerated ones such as ``DD(AVX2)". The corresponding results of the same precision mpmath (MP) computation are also listed in it.

\begin{table}[htbp]\small
\begin{center}
\caption{Computational time of Direct methods}\label{table:lubench}
\begin{tabular}{|c|c|c|}
\multicolumn{3}{c|}{Corei9: $n=2000$}  \\ \hline
Prec. Direct Methods & Time(s) & Max.RE \\ \hline
DD & 12.0 & 1.0E-01  \\
DD(AVX2) & 4.5 & 1.7E-01  \\
MP(106bits) & 17365.3 & 2.7E-05  \\ \hline
TD & 83.1 & 9.4E-18 \\
TD(AVX2) & 42.6 & 4.6E-18  \\
MP(159bits) & 17198.4 & 1.2E-18 \\ \hline
QD & 297.6 & 2.7E-34 \\
QD(AVX2) & 64.8 & 1.4E-34 \\
MP(212bits) & 17090.9 & 1.8E-34 \\ \hline
\end{tabular}
\begin{tabular}{|c|c|}
\multicolumn{2}{|c}{EPYC: $n=2000$}  \\ \hline
Time(s) & Max.RE  \\ \hline
15.4 & 1.0E-01 \\
4.4 & 1.7E-01 \\
26663.5 & 2.7E-05 \\ \hline
99.9 & 9.4E-18 \\
56.1 & 4.6E-18 \\
26645.9 & 1.2E-18 \\ \hline
380.8 & 2.7E-34 \\
102.3 & 1.4E-34 \\
26832.9 & 1.8E-34 \\ \hline
\end{tabular}
\end{center}
\end{table}

For DD precision direct method, the same precision mpmath direct method (lu\_solve function) can obtained more accurate numerical solutions. On the other hand, TD and QD precision direct methods can obtain the same level of accuracy as mpmath direct methods. Computational time of accelerated direct methods with AVX2 can be reduced drastically compared with mpmath. Direct methods of mpmath is 262 to 6028 times slower than our non-SIMDized and SIMDized direct methods.

From these results, we can expect to obtain good performance of DD, TD and QD-mpmath mixed precision iterative refinement on Python. \tablename\ \ref{table:iterrefbench} shows all results of the mixed precision iterative refinement with 424bits mpmath computation. It has computational time (Time(s)), maximum relative errors of numerical solutions(Max.RE) and the number of iterations (\#Iter.), and also shows the results of 424 direct methods of mpmath. Stopping condition in process of iterations is set as (\ref{eqn:iterref_stop_cond}), where $\mathbf{x}_k$ is the $k$-th approximation and $\mathbf{r}_k := \mathbf{b} - A\mathbf{x}_k$.
\begin{equation}
	\|\mathbf{r}_k\|_2 \leq \sqrt{n} \cdot 10^{-100} \cdot \|A\|_F\cdot \|\mathbf{x}_k\|_2 \label{eqn:iterref_stop_cond}
\end{equation}

\begin{table}[htbp]\small
\begin{center}
\caption{Computational time of mixed precision iterative refinement methods}\label{table:iterrefbench}
\begin{tabular}{|c|c|c|c|}
\multicolumn{4}{c|}{Corei9: $n=2000$}  \\ \hline
Prec. Iter.refinement & Time(s) & Max.RE & \#Iter. \\ \hline
DD+MP(424bits)      & 190.7 & 2.2E-78 & 15 \\
DD+MP(AVX2) & 177.0 & 2.7E-77 & 15 \\ \hline
TD+MP       & 171.0 & 1.0E-80 & 2 \\
TD+MP(AVX2) & 131.3 & 8.8E-80 & 2 \\ \hline
QD+MP       & 409.8 & 7.5E-99 & 1 \\
QD+MP(AVX2) & 176.5 & 1.8E-98 & 1 \\ \hline
MP Direct(424bits)   & 17940.8 & 8.8E-99 & N/A \\ \hline
\end{tabular}
\begin{tabular}{|c|c|c|}
\multicolumn{3}{|c}{EPYC: $n=2000$}  \\ \hline
 Time(s) & Max.RE & \#Iter. \\ \hline
 294.2 & 2.2E-78          & 16 \\
 279.4 & 2.7E-77          & 15 \\ \hline
 241.0 & 1.0E-80          & 2 \\
 200.0 & 8.8E-80          & 2 \\ \hline
 564.6 & 7.5E-99          & 1 \\
 287.9 & 1.8E-98          & 1 \\ \hline
 27700.3 & 8.8E-99        & N/A \\ \hline
\end{tabular}
\end{center}
\end{table}

It can be seen that the maximum relative errors are different for mixed precision iterative refinements with different computational precision, but normally accurate approximations are obtained for all precision. Mixed precision iterative refinements combined with multi-component-type MPF direct methods are 49 to 139 times faster than 424bits direct method of mpmath. SIMDized DD, TD, and QD precision direct methods can accelerate iterative refinement process, especially for QD-mpmath case.

As shown in \tablename\ \ref{table:iterrefbench}, the minimum computational time can be obtained when TD-mpmath iterative refinement is used. Therefore, our accelerated MPF direct methods can optimize the process to solve linear system of equations by selecting appropriate precision of direct method according to the condition of the equations and the computing environment.


%
\section{Future works}
Our future plan is to provide a parallelized version of BNCmatmul in OpenMP on Python environment in order to expand the variety of MPF calculations as part of Python eco-system.


%
\section*{ACKNOWLEDGEMENTS}

This study is supported by JSPS Kakenhi (JP20K11843) and Shizuoka Institute of Science Technology.
%


\end{document}